\documentclass[aps,prd,twocolumn,tightelines, superscriptaddress,nofootinbib]{revtex4}
\usepackage{graphicx}
\usepackage{epsfig}
\usepackage{bm}
\usepackage{latexsym,amssymb,amsmath,amsfonts,amssymb,txfonts,pxfonts,wasysym,float}
\usepackage{mathrsfs}
\usepackage{color}
\usepackage{lettrine}
\usepackage{lipsum}
\usepackage{enumitem}

\newcommand{\postscript}[2]{\setlength{\epsfxsize}{#2\hsize}
   \centerline{\epsfbox{#1}}}

\newcommand\be{\begin{equation}}
\newcommand\ee{\end{equation}}

\newcommand\bea{\begin{eqnarray}}
\newcommand\eea{\end{eqnarray}}

\usepackage[usenames,dvipsnames]{xcolor}
\definecolor{orange}{cmyk}{0,0.5,1,0}
\definecolor{rossoCP3}{cmyk}{0,.88,.77,.40}
\definecolor{graa}{rgb}{0.8,0.8,0.8}
\definecolor{blaa}{rgb}{0.2,0.2,0.6}

\begin{document}

\title{\color{rossoCP3} {A Physics Modeling Study of SARS-CoV-2 Transport in Air}}    

\author{\bf Luis A. Anchordoqui}
\email{luis.anchordoqui@lehman.cuny.edu}
\affiliation{Department of Physics and Astronomy,  Lehman College, City University of
  New York, NY 10468, USA
}

\affiliation{Department of Physics,
 Graduate Center, City University
  of New York,  NY 10016, USA
}

\author{\bf James B. Dent }
 \email{jbdent@shsu.edu}
\affiliation{Department of Physics, Sam Houston State University, Huntsville, TX 77341, USA}

\author{\bf Thomas J. Weiler}
\email{tom.weiler@vanderbilt.edu}
\affiliation{Department of Physics and Astronomy, Vanderbilt University, Nashville, TN 37235, USA}

\date{August 2020}

\begin{abstract}
\noindent {\it General Idea:} Health threat from SARS-CoV-2 airborne infection
has become a public emergency of international concern. During the
ongoing coronavirus pandemic, people have been advised by the Centers
for Disease Control and Prevention to maintain social distancing of at
least 2~m to limit the risk of exposure to the
coronavirus. Experimental data, however, show that infected aerosols and
droplets trapped inside a turbulent puff cloud can travel 7 to 8~m. We carry out a physics modeling study for SARS-CoV-2
transport in air.\\
{\it Methodology:} We propose a nuclear physics analogy-based modeling
of the complex gas cloud and its payload of pathogen-virions. The cloud
is modeled as a spherical puff of hot moist air 
(with mucosalivary filaments), which remains coherent in a volume that varies from 0.00025 to
0.0025~${\rm m}^3$. The puff propagates scattering off
the air molecules. We estimate the puff effective stopping range
adapting the high-energy physics model that describes the slow down of $\alpha$-particles
(in matter) via interactions with the electron cloud.\\
{\it Research Findings:} We show that the stopping range is
proportional to the product of the puff's diameter and its density. We use our puff model to determine the
average density of the buoyant fluid in the turbulent cloud. A fit to
the experimental data yields  $1.8 < \rho_P/\rho_{\rm air} < 4.0$, where
$\rho_P$ and $\rho_{\rm air}$ are the average density of the puff and
the air. We  demonstrate
that temperature variation could cause an ${\cal O}(\pm 8\%)$ effect in
the puff stopping range for extreme ambient cold or warmth. We also 
demonstrate that aerosols and droplets can remain suspended for
hours 
in the air. Therefore, once the puff slows down sufficiently,
and its coherence is lost, the eventual spreading of the infected
aerosols becomes dependent on the ambient air currents and
turbulence. \\
{\it Practical Implications:} Viral transmission pathways have
profound implications for public safety. Our study forewarns a health
threat of SARS-CoV-2 airborne infection in indoor spaces. We argue in favor of implementing additional precautions to the recommended 2~m social distancing, e.g. wearing a face mask when we are out in public.
\end{abstract}
\maketitle

\section{Introduction}

The current outbreak of the respiratory disease identified as COVID-19 is caused by
the severe acute respiratory syndrome coronavirus 2, shortened to
SARS-CoV-2~\cite{Huang,Zhou,Zhu,Tang}. The outbreak was first reported in December 2019, and has become a worldwide pandemic with over 10 million cases as of 1 July 2020.
SARS-CoV-2 have been confirmed worldwide 
and so the outbreak has been declared 
a global pandemic by the World Health Organization. The pandemic has spread around the globe to almost every region, with only a handful of the World Health Organization's member states not yet reporting cases. Most of these states are small island nations in the Pacific Ocean, including Vanuatu, Tuvalu, Samoa, and Palau. 

The coronavirus can spread
from person-to-person in an efficient and sustained way by coughing
and sneezing. The virus can spread from seemingly healthy carriers or people who had not yet developed symptoms~\cite{Rothe}. To understand and prevent the
spread of the virus, it is important to estimate the probability of
airborne transmission as aerosolization with particles potentially
containing the virus. Before proceeding, we pause to note that herein we follow the convention of the World Health Organization and refer to 
particles which are $\agt 5~\mu$m diameter as droplets and those $\alt 5~\mu$m as aerosols or droplet nuclei~\cite{Shiu}.

There are various experimental measurements suggesting that SARS-CoV-2 may have the potential to be transmitted through aerosols; see e.g.~\cite{Ong,Santarpia,Liu,Cai,Guo}. Indeed, laboratory-generated aerosols with SARS-CoV-2 were found to keep a replicable virus in cell culture throughout the 3 hours of aerosol testing~\cite{Doremalen}. Of course these laboratory-generated aerosols may not be exactly analogous to human exhaled droplet nuclei, but they helped in establishing that the survival times of SARS-CoV-2 depend on its environment, including survival times of: up to 72 hours on plastics, up to 48 hours on stainless steel, up to 24 hours on cardboard, up to 4 hours on copper, and in air for 3 to 4 hours~\cite{Doremalen}. On first glimpse this finding is surprising, as one would expect that the properties of air that
degrade the SARS-CoV-2 exterior should abate at roughly half that time if it were adhered to a surface (i.e. at least half
the solid angle is mostly exposed to air). However, the laboratory-generated aerosols have shown that a precise description of SARS-CoV-2 main characteristics  requires more complex systems in which the virus would be chemisorbed
by some surfaces and repelled by the others. More concretely, the survival
probability of the virus is associated with the surface energies of the
various materials that can reduce the solid angle exposed to air molecule collisions. These properties can lead to remarkable differences , for example that between copper and
stainless steel. Despite the fact both are metals, copper causes destruction of the virus much more rapidly than does stainless steel.

The number of virions needed for infection is yet unknown. However, it is known that viral load differs considerably between SARS-CoV and SARS-CoV-2~\cite{Wolfel}. A study of the variance of viral loads in patients of different ages found no significant difference between any pair of age categories including children~\cite{Jones}.

Beyond a shadow of a doubt, a major question of this pandemic has been how far would be far enough to elude droplets and to diffuse droplet nuclei if a person nearby is coughing
or sneezing. The rule of thumb for this pandemic has been a 2~m separation. Nevertheless, this has never been a magic number that guarantees
complete protection. Indeed, experiment shows that: {\it (i)}~respiratory particles emitted  during a sneeze or cough are initially
transported as a turbulent cloud that consists of hot and moist
exhaled air and mucosalivary filaments; {\it (ii)}~aerosols and
small droplets trapped in the turbulent puff cloud could propagate 7
to
8~m~\cite{Bourouiba_sneeze,Bourouiba,Scharfman,Bourouiba2}. Moreover, once the cloud slows down sufficiently, and its coherence is lost, the eventual spreading of the infected aerosols becomes dependent on the ambient air currents and turbulence~\cite{Anchordoqui}. In this paper we provide new guidance to address this question by introducing a physics model for SARS-CoV-2 transport in air.

To develop some sense for the orders of magnitude involved, we begin
by reviewing the experimental data. A survey of 26 analyses reporting
particle sizes generated from breathing, coughing, sneezing and
talking indicates that healthy individuals generate particles with
sizes in the range $0.01 \alt D_V/\mu{\rm m} \alt 500$, whereas
individuals with infections produce particles in the range $0.05 \alt
D_V/\mu{\rm m} \alt 500$, where $D_V$ is the diameter of a respiratory
particle (droplet or droplet nucleus)  containing the
virus~\cite{Gralton}. The majority of the particles containing the virus have outlet velocities in the range $10 \alt v_{V,0}/({\rm m/s}) \alt 30$~\cite{Xie,Wei,Bourouiba2}. 
Up to $10^{4.6}$ particles are expelled at an initial velocity of 30~m/s
during a sneeze, and a cough can generate approximately $10^{3.5}$
particles with outlet velocities of 20~m/s~\cite{Cole}. 97\% of coughed particles have sizes $0.5 \alt
D_V/\mu{\rm m} \alt 12$, and the primary size distribution is within the
range $1 \alt
D_V/\mu{\rm m} \alt 2$~\cite{Duguid,Yang}. 
The evaporation rate of the respiratory particles depends on the exposed surface area,
$A \sim \pi D_V^2$, while the particle's volume scales as $V\sim \pi
D_V^3/6$. Therefore, the ratio of area to volume is $A/V \propto
1/D_V$, and it is the smallest droplets that will live the
longest. 

The layout of the paper is as follows. In Sec.~\ref{sec:2} we review
the generalities of  aerodynamic drag force and estimate the terminal speed of
aerosols and droplets. In Sec.~\ref{sec:3}   we model the elastic
scattering of the turbulent cloud with the air molecules and
estimate the puff stopping range assuming standard ambient temperature and
pressure conditions. After that, we use our puff model to determine
the average density of the buoyant fluid in the turbulent cloud. The paper wraps up with some conclusions presented in Sec.~\ref{sec:4}.

\section{Terminal Speed}
\label{sec:2}

When a particle propagates through the air, the surrounding air
molecules have a tendency to resist its motion. This resisting force
is known as  the aerodynamic drag force. For a 
spherical particle, the aerodynamic drag force is given by
\begin{equation}
\textbf{F}_d = 3 \pi \ \eta_{\rm air} \ D_V \  \textbf{v}_V \frac{1}{\varkappa} \, .
\label{drag}
\end{equation}
where $\eta_{\rm air} \simeq 1.8 \times 10^{-5}~{\rm kg/(m\cdot s)}$ is the dynamic
viscosity of air and $\textbf{v}_V$ is the virus velocity
vector. Eq.(\ref{drag}) is the well-known Stokes' law, with the Cunningham
slip correction factor $\varkappa$; see Appendix for details. Stokes' law assumes that the relative velocity of a carrier gas at a
particle's surface is zero; this assumption does not hold for small
particles. The slip correction factor should be applied to Stokes' law
for particles smaller than $10~\mu {\rm m}$. 

The particle Reynolds number, 
\begin{equation}
{\cal R} = \frac{D_V \ v_V \ \rho_{\rm air}}{\eta_{\rm air}} \,,
\end{equation}
is a dimensionless quantity which represents the ratio of inertial
forces to viscous forces, where $\rho_{\rm air} \simeq 1.2~{\rm
  kg}/{\rm m}^3$ is the air density at a temperature of 20$^{\circ}$ C (293 K). For ${\cal R} < 1$, the inertial
forces can be neglected. The drag calculated by Eq.(\ref{drag}) has an
error of about 12\% at ${\cal R} \approx 1$. The error decreases with decreasing particle Reynolds number.

\begin{figure}[tb]
\postscript{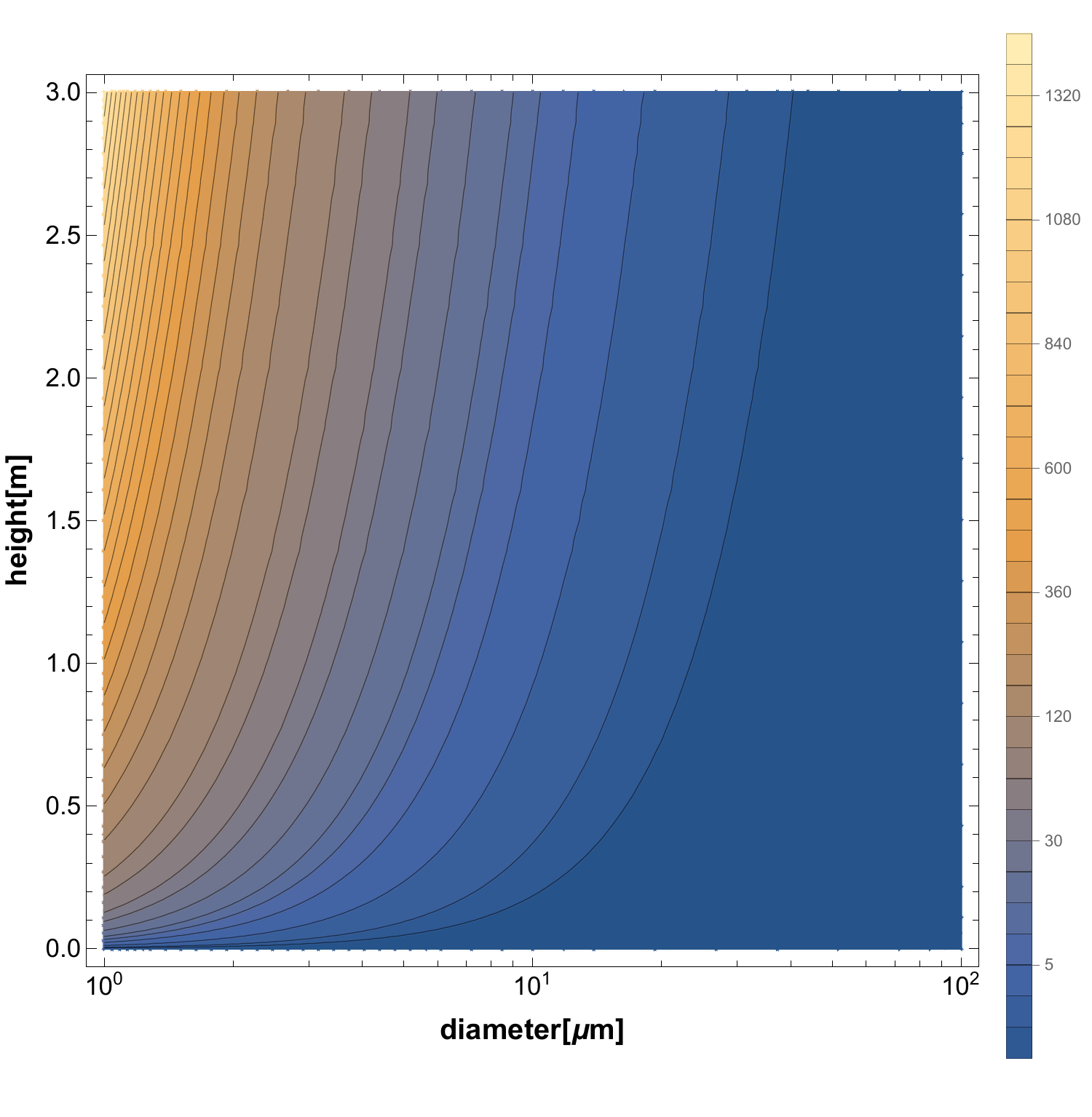}{0.9}
\caption{Contours of the time $t_f$ in minutes in the
  $h-D_V$ plane. \label{fig:1}}
\end{figure}

For the case at hand, ${\cal R} >1$. In the vertical direction, the
upward component of the aerodynamic drag force $F_{d,\perp}$ is
counterbalanced by the excess of the gravitational attraction over the
air buoyancy force 
\begin{equation}
  F_g = \frac{1}{6} \ \pi \ D_V^3 \ (\rho_{\rm H_2O} - \rho_{\rm air})
  \ g \, ,
\end{equation}
where $\rho_{\rm H_2O} \simeq 997~{\rm kg}/{\rm m}^3$ and $g \simeq
9.8~{\rm m/s}^2$ is the acceleration of gravity. Since $\rho_{\rm air}
\ll \rho_{\rm H_2O}$ the air buoyancy force
becomes negligible, and so $F_g \approx M_V g$, with $M_V$ the aerosol
mass. When the upward
aerodynamic drag force
equals the gravitational attraction the droplet reaches mechanical
equilibrium and starts falling with a terminal speed
\begin{equation}
v_{V,f,\perp} \approx \frac{M_V \ g \ \varkappa}{3 \pi \ \eta \ D_V} \, .
\end{equation}
The terminal speed is $\propto D_V^2$ (due to the diameter dependence of the mass), and hence larger droplets
would have larger terminal velocities thereby reaching the ground
faster. The terminal speed for various particle sizes is given in
Table~\ref{tabla1}. The time $t_f$ it will take the virus to fall to the
ground is simply given by the distance to the ground divided by
$v_{V,f,\perp}$. For an initial height, $h \sim 2~{\rm m}$, we find
that for $D_V = 2~\mu{\rm m}$,
\be
t_f = \frac{h}{v_{V,f,\perp}} \sim 4~{\rm hr} \, .
\ee
The time scale as a function of the droplet size and heigth is shown in Fig.~\ref{fig:1}.

The aerodynamic drag force holds for rigid spherical particles moving
at constant velocity relative to the gas flow. To determine the
stopping range, in the next section we model the elastic scattering of
the turbulent puff cloud with the air molecules.

\begin{table}
\caption{Cunningham slip correction factor and terminal speed. \label{tabla1}}
  \begin{tabular}{ccc}
    \hline
    \hline
     $D_V~(\mu{\rm m})$ & $\varkappa$ & $v_{V,f,\perp}~({\rm m/s})$   \\ 
\hline
   $\phantom{1}0.001$ & $215.3$  & $6.51 \times 10^{-9}$ \\
    $\phantom{1}0.010$ & $\phantom{2}22.05$ & $6.67 \times 10^{-8}$ \\
   $\phantom{1}0.100$ &  $\phantom{22}2.851$  & $8.62 \times 10^{-7}$ \\
   $\phantom{1}0.500$ &  $\phantom{22}1.327$  & $1.00 \times 10^{-5}$ \\
   $\phantom{1}1.000$ & $\phantom{22}1.163$  & $3.52 \times 10^{-5}$ \\
   $\phantom{1}1.500$ & $\phantom{22}1.109$  & $7.54 \times 10^{-5}$ \\
   $\phantom{1}2.000$ & $\phantom{22}1.081$  & $1.31 \times 10^{-4}$ \\
   $\phantom{1}3.000$ & $\phantom{22}1.054$  & $2.87 \times 10^{-4}$ \\
   $\phantom{1}5.000$ & $\phantom{22}1.033$  & $7.81 \times 10^{-4}$ \\
    $\phantom{1}7.000$ & $\phantom{22}1.023$ & $1.52 \times 10^{-3}$ \\
    ~~~~~~~~~$10.000$~~~~~~~~~ & ~~~~~~~~~$\phantom{22}1.016$~~~~~~~~~  & ~~~~~~~~~$3.07 \times 10^{-3}$~~~~~~~~~ \\
    \hline
    \hline
  \end{tabular}
\end{table}

\section{Stopping Range}
\label{sec:3}

Respiratory particles of saliva and mucus are expelled together
with a warm and humid air, which generates a convective current. The
aerosols and droplets are initially transported as part of a 
coherent gas puff of buoyant fluid.  The ejected puff of air remains
coherent in a volume that varies from $0.00025$ to $0.0025~{\rm m}^3$~\cite{Balachandar}.
This corresponds to a puff size $0.78 \lesssim D_P/{\rm m} \lesssim
1.68$, where follwoing~\cite{Balachandar} we have taken an entrainment
coefficient~\cite{Morton} of $\alpha = 0.1$. 
The puff  is ejected with $1 \lesssim v_{V,0,\parallel}/{\rm
  (m/s)} \lesssim 10$~\cite{Balachandar}. 
The turbulent puff cloud consists of an admixture of moist exhaled air and
mucosalivary filaments. Next, in line with our stated plan, we use the
experimental data to calculate the range of the average
density of the buoyant fluid in the turbulent cloud.

The mass ratio of the average air molecule compared to the aerosol, $m_{\rm{air}}/ M_V$, is roughly $10^{-12} $
(since the size of the aerosol and the mass for its chief constituent,
H$_2$O, compared to the air molecule are $10^4$ and $10^{3}$), though
there is an obvious variation with aerosol size at constant
density. If we consider instead the mass inside the puff $M_P$ the ratio $R \equiv
m_{\rm{air}}/ M_{P}$ is even smaller. Due to the enormous mass ratio, the virions inside
the puff will not undergo large angular deflections, so we will treat the virions as having the same direction for its initial and final velocities (since we are looking at a stopping distance, this is a reasonable assumption). Starting with the non-relativistic one-dimensional equation for the virus velocity $\beta$ we have in the lowest nontrivial order (in $R \ll 1$) and any frame
\bea       
\left(
\begin{array}{c}
\beta_1\\
v_{{\rm air},f}
\end{array}
\right)
= \mathbb{M} \left(\begin{array}{c} \beta_0 \\ v_{{\rm air},0} \end{array} \right) \,,
\eea
where the matrix $\mathbb{M}$ is derived by imposing conservation of energy and momentum, and
is given by
\bea
\mathbb{M} = \left(
\begin{array}{cc}
~~1-2R~~  & ~~2R~~ \\
~~2~~ &   ~~ -1~~
\end{array}
\right) \,,
\eea
with $\beta_0 = v_{V,0,\parallel}$, and $v_{{\rm air},0}$ and $v_{{\rm air},f}$ the initial and final velocities of the air molecule, respectively.  
As the velocity $ \beta $ falls with each interaction, the velocity
loss remains constant; the target particle is a new air molecule at
each interaction.

Though individual air molecules are traveling at an average speed of a few hundred meters per
second, throughout we assume the medium to be stationary. In analogy
with the description of the slowing down of alpha particles in matter (which assumes the
electronic cloud is at rest),  we can describe the
scattering of the puff in the frame in which the air molecule is at rest, i.e.,
$v_{{\rm air},0} = 0$ (in essence, adopting a stationary medium on average). The stopping power is given by the velocity-loss equation
\bea
 d\beta/dx = \Delta\beta/\lambda^V_{\rm mfp} =
2R \beta/\lambda^V_{\rm mfp}  \,,
\eea
with solution 
$ \ln\beta= ( 2R / \lambda^V_{\rm mfp} ) \int dx $. Finally, we have
for the stopping distance
\bea 
L=\lambda^V_{\rm mfp} \ \frac{1}{2R} \ \ln\left(\frac{\beta_0}{\beta_f} \right) \,,
\label{stopdis}
\eea
with $\beta_f \equiv v_{V,f,\parallel}$. Note that $L/\lambda_{\rm mfp}^V$ is not only the number of mean free paths traversed by the fiducial virus, but is also the number of interactions of the virus with air molecules; of course, there is a one-to-one correlation between the number of mean free paths traveled and interactions.          

Since $\beta$ is homogeneous and the mass ratio $R$ is a constant for
a given puff size $D_P$, we
have the above simple equation. The mass ratio $R$ is very small, and
$(2R)^{-1}$ is correspondingly very large. There are a tremendous number of mean free paths/interactions involved as the virions bowling ball rolls over the air molecule.

Finally, we must calculate   
$ \lambda^V_{\rm mfp} = 1/ (n_{\rm air} \sigma) $. The air molecules
act collectively as a fluid, so the volume $V$ over the air density is
given by the ideal gas law as $ k_{\rm B}T/P $, where $P$ is the
pressure, $T$ the temperature, and $k_{\rm B}$ is the Boltzmann
constant. We assume a contact interaction equal to the cross-sectional hard-sphere size of the puff, 
i.e. $ \sigma = \pi (D_P/2)^2 $. Substituting into Eq.(\ref{stopdis}) we
obtain the final result for the stopping distance
\begin{equation}
L= \frac{k_{\rm B}T}{P}  \ \frac{1}{\pi (D_P/2)^2} \frac{1}{2R} \
\ln \left(\frac{\beta_{0}}{\beta_{f}} \right) \, .
\label{buga}
\end{equation}
We take the sneeze or cough which causes the droplets expulsion to be
at a standard ambient air pressure of $P = 101$~kPa and
a temperature of $T \sim 293~{\rm K}$. It is important to stress that {\it temperature variation could cause an $\mathcal{O}(\lesssim \pm8 \%)$ effect in $L$ for extreme ambient cold or warmth}. We now proceed to fit the experimental data. For $L \sim 8~{\rm m}$ and
taking  $v_{V,f,\parallel} \sim 3~{\rm mm/s}$~\cite{Bourouiba},
 we obtain $1.8 <
\rho_P/\rho_{\rm air} < 4.0$ for $0.78 \lesssim D_P/{\rm m} \lesssim
1.68$, 
where
$\rho_P$ is the average density of the fluid in the puff. 

A point worth noting at this juncture is that our model provides an
effective description of the turbulent puff cloud. Note that
independently of their size and their initial velocity all respiratory
particles in the cloud experience both gravitational settling and
evaporation. Aerosols and droplets of all sizes are subject to
continuous settling, but those with settling speed smaller than the
fluctuating velocity of the surrounding puff would remain trapped
longer within the puff. Actually, because of evaporation the water
content of the respiratory particles is monotonically decreasing. At
the point of almost complete evaporation the settling velocity of the
aerosols is sufficiently small that they can remain trapped in the puff
and get advected by ambient air currents and dispersed by ambient
turbulence. The size of the puff then continuously grows in
time~\cite{Balachandar}. Our result can equivalently be interpreted in
terms of the effective coherence length of the turbulent cloud
assuming $\rho_P \sim \rho_{\rm air}$. The effective size of the puff
and its effective density are entangled in Eq.~(\ref{buga}). Numerical
simulations show that during propagation
the puff edge grows $\propto t^{1/4}$~\cite{Bourouiba}. After a 100~s the puff would grow by
a factor of 3 (see Fig.~7 in~\cite{Balachandar}), in agreement with
our analytical estimates. In
closing, we note that if we ignore the motion of the air puff carrying
the aerosols, as in the analysis of~\cite{Wells}, it is
straightforward to see substituting $R$ by $m_{\rm air}/M_V \sim
10^{-12}$ into Eq.~(\ref{buga}) that the individual aerosols would not travel
more than a few cm away from the exhaler, even under conditions of
fast ejections, such as in a sneeze. This emphasizes the relevance of
incorporating the complete multiphase flow physics in the modeling of
respiratory emissions when ascertaining the risk of SARS-CoV-2 airborne
infection.

\section{Conclusions}
\label{sec:4}
  
We have carried out a physics modeling study for SARS-CoV-2 transport
in air. We have developed a nuclear physics analogy-based modeling of
the complex gas cloud and its payload of pathogen-virions. Using our
puff model we estimated the average density of the fluid in the
turbulent cloud is in the range  $1.8 < \rho_P/\rho_{\rm air} < 4.0$. We have also shown that aerosols and droplets can remain suspended
for hours in the air. Therefore, once the puff slows down sufficiently,
and its coherence is lost, the eventual spreading of the infected
aerosols becomes dependent on the ambient air currents and
turbulence. De facto, as it was first pointed out in~\cite{Anchordoqui} and later developed in~\cite{Augenbraun,Evans} airflow
conditions strongly influence the distribution of viral particles in
indoor spaces, cultivating a health threat from COVID-19 airborne
infection.

Altogether, it seems reasonable to adopt additional infection-control
measures for airborne transmission in high-risk settings, such as the
use of face masks when in public. 
If the results of this study - $t_f$ of ${\cal O} ({\rm hr})$ for aerosols, for example - are borne out by experiment, then these findings should be taken into account in policy decisions going forward as we continue to grapple with this pandemic.

\section*{Appendix}

There are important considerations in the development of Stokes' law,
including the hypothesis that the gas at particle
surface has zero velocity relative to the particle. This hypothesis
holds well when the diameter of the particle is much larger than the mean
free path of gas molecules. The mean free path $\lambda_{\rm mfp}^{\rm
  air}$ is the average distance
traveled by a gas molecule between two successive collisions. In
analyses of the interaction between gas molecules and particles, it is
convenient to use the Knudsen number ${\rm Kn} = 2 \lambda_{\rm mfp}^{\rm
  air}/D_V$, a dimensionless number
defined as the ratio of the mean free path to particle radius. For ${\rm Kn}
\agt 1$, the drag force is smaller than
predicted by Stokes' law. Conventionally this condition is described as
a result of slip on the particle surface. The so-called slip
correction is estimated to be~\cite{Crowder}
\begin{equation}
\varkappa = 1 + {\rm Kn} \left[1.257 + 0.4 \ \exp(-1.1/{\rm Kn}) \right] \, .
\end{equation}
In our calculations we take 
\bea
\lambda_{\rm mfp}^{\rm air} = \frac{\eta_{\rm air}}{\rho_{\rm
    air}}\left(\frac{\pi m_{\rm air}}{2 \, k_{\rm B} \, T}\right)^{1/2} \,,
\eea
where $k_{\rm B}$ is the Boltzmann constant, $T$ is the temperature in
Kelvin, and the density of air is given by
\bea
\rho_{\rm air} = \frac{P}{R_gT} \,,
\eea
with $P = 101$ kPa, and where $R_g = 287.058$~J/(kg~$\cdot$~K) is the ideal gas constant. The molar mass of air is
$m_{\rm mol} = 29\; {\rm g}/{\rm mol}$, which leads to $m_{\rm air} =  4.8\times10^{-26}\;{\rm kg}/{\rm molecule}$.\\

\noindent{\bf Funding/Support:} The research of L.A.A. is supported by the U.S. National Science
Foundation (NSF Grant PHY-1620661). J.B.D. acknowledges support from
the National Science Foundation under Grant No. NSF PHY182080. The
work of T.J.W. was supported in part by the U.S. Department of Energy
(DoE grant No. DE-SC0011981).

\noindent {\bf Role of the Funder/Sponsor:} The sponsors had no role in
the preparation, review or approval of the manuscript and decision to
submit the manuscript for publication. Any opinions,
findings, and conclusions or recommendations expressed in this
article are those of the authors and do not necessarily reflect the
views of the NSF or DOE.

\noindent{\bf Declaration of Competing Interest:} The authors declare that they have no known competing financial interests or personal relationships that could have appeared to influence the work reported in this paper.

\noindent{\bf  Ethical Approval:} The manuscript does not contain experiments on animals and humans; hence ethical permission not required.

\end{document}